\documentclass[letterpaper, 10 pt, conference]{ieeeconf}  

\IEEEoverridecommandlockouts                              
\usepackage{graphics} 
\usepackage{epsfig} 
\usepackage{times} 
\usepackage{amsmath} 
\usepackage{amssymb}  

\DeclareMathOperator{\tr}{tr}
\newcommand{\E}{\mathbb{E}}
\newcommand{\KL}{\textrm{KL}}
\newtheorem{theorem}{Theorem}
\newtheorem{definition}{Definition}

\newtheorem{proposition}{Proposition}
\newtheorem{corollary}{Corollary}

\title{\LARGE \bf
Sample Complexity Lower Bounds for Linear System Identification
}

\author{Yassir Jedra and Alexandre Proutiere
\thanks{Y. Jedra and A. Proutiere are with the Division of Decision and Control Systems, School of Electrical Engineering and Computer Science, Royal institute of Technology (KTH), Stockholm, Sweden. Emails: \{{\it jedra@kth.se, alepro@kth.se}\}.}
}

\begin{document}

\maketitle
\thispagestyle{empty}
\pagestyle{empty}

\begin{abstract}
This paper establishes problem-specific sample complexity lower bounds for linear system identification problems. The sample complexity is defined in the PAC framework: it corresponds to the time it takes to identify the system parameters with prescribed accuracy and confidence levels. By problem-specific, we mean that the lower bound explicitly depends on the system to be identified (which contrasts with minimax lower bounds), and hence really captures the identification hardness specific to the system. We consider both uncontrolled and controlled systems. For uncontrolled systems, the lower bounds are valid for {\it any} linear system, stable or not, and only depend of the system finite-time controllability gramian. A simplified lower bound depending on the spectrum of the system only is also derived. In view of recent finite-time analysis of classical estimation methods (e.g. ordinary least squares), our sample complexity lower bounds are tight for many systems. For controlled systems, our lower bounds are not as explicit as in the case of uncontrolled systems, but could well provide interesting insights into the design of control policy with minimal sample complexity. 
\end{abstract}
\section{Introduction}

System identification is concerned with the design of procedures to estimate the unknown parameters of a dynamical system. Linear Time Invariant (LTI) systems constitute an important class of models investigated in system identification, and arise in many fields such us finance, biology, robotics and other engineering and control applications \cite{Ljung:1986:SIT:21413}. In the past, theoretical results related to the identification of LTI systems have been mainly concerned with guarantees on the asymptotic convergence properties of particular estimation schemes \cite{Goodwin:1977:DSI}, for example the ordinary least squares method \cite{Ljung:c1}, maximum likelihood and prediction error methods \cite{Ljung:c2}. In contrast, finite-time properties of estimation schemes for LTI systems had been less studied. Such an analysis is motivated for example in linear quadratic regulator problems with unknown dynamics. There, the overall performance of the control policy depends on the identification errors \cite{dean:2017, arora2018towards}. Therefore, a deeper understanding of the finite-time behavior of estimation schemes constitutes an important objective.     

In fact, the finite-time behavior of LTI systems has received substantial attention in both the control and machine learning communities. Most of these recent efforts focus on the finite-time properties of classical estimation schemes (mainly ordinary least squares methods) for observable states \cite{rantzer2018,faradonbeh:2018:c2, simchowitz:2018a, sarkar:2018, oymak:2018, sarkar:2019} or unobservable states \cite{oymak:2018, sarkar:2019}. The results derived there heavily rely recent results in statistical learning theory, random matrix theory \cite{mendelson:2014, vershynin:2012}, and self-normalizing processes \cite{pena2008self}. 

Some of the aforementioned recent works also provide information-theoretical arguments to derive lower bounds on the number of observations required to identify the system parameters with a prescribed level of accuracy. These sample complexity lower bounds are typically derived in a minimax framework (they characterize the minimal number of observations required for the {\it worst} system), and are restricted to specific subclasses of systems \cite{simchowitz:2018a}. These limitations are shared by earlier studies on the sample complexity of LTI system identification \cite{weyerLTI:1999, raginsky:2010}. Refer to Section \ref{sec:related} for a more exhaustive survey. 

In this paper, we derive sample complexity lower bounds for LTI system identification. Our bounds are {\it problem-specific} in the sense that they depend on the parameters of the system explicitly, which contrasts with minimax lower bounds. This dependence reveals the hardness of the parameter estimation problem specific to each particular system. Our bounds are valid for {\it any} LTI system, stable or not. The derivation of our bounds relies on the use of change-of-measure arguments, originally used in the analysis of fundamental limits in stochastic optimization problems such as bandit optimization \cite{lai1985}. These arguments have been instrumental in problems with finite sets of decisions (e.g. to derive the sample complexity of identifying the best arm among a finite set in bandit optimization problems). To get sample complexity lower bounds in LTI systems, we adapt these arguments to problems with continuous sets of decisions (the estimated system parameters are real matrices). Importantly, we note that change-of-arguments typically lead to tight fundamental bounds, and we believe our bounds are tight (they actually match the finite-time performance of some of classical estimation procedures for many systems).   

We first treat the case of uncontrolled LTI systems with observable states: $x_{t+1}=Ax_t+w_t$ (with Gaussian noise $w_t$). We establish (see Theorem \ref{th:uncontrolled}) that the number of observations $\tau_A$ required to derive an estimator $\hat{A}$ of $A$ with $\mathbb{P}[\| \hat{A}-A\|_F\le \epsilon]\ge 1-\delta$ (for some $\epsilon>0$ and $\delta\in (0,1)$) satisfies $\lambda_{\min}\Big(\sum_{s=1}^{\tau_A -1} \Gamma_{s-1}(A) \Big) \ge \frac{1}{2\epsilon^2}\log(\frac{1}{2.4 \delta})$, where $\Gamma_{s}[A]=\sum_{k=0}^sA^k(A^k)^\top$ is the finite-time controllability gramian of the system\footnote{$\|\cdot\|_F$ is the Frobenius norm,and  $\lambda_{\min}(S)$ is the minimal eigenvalue of symmetric matrix $S$.}. We further simplify this condition, and show that $\tau_A$ also satisfies $\phi_{|\lambda_d(A)|}(\tau_A) \ge \frac{1}{2\epsilon^2}\log(\frac{1}{2.4 \delta})$, where $\lambda_d(A)$ denotes the complex eigenvalue of $A$ with smallest amplitude, and where $\phi_a(\cdot)$ is an explicit family of mappings. We then extend the results to controlled LTI systems. The lower bounds are not as explicit as in the case of uncontrolled systems, but could provide interesting insights into the design of control policy with minimal sample complexity. We specify the bounds in the case of scalar systems.

\section{Related work}\label{sec:related}

Some of the recent papers towards a finite-time analysis of LTI systems provide sample complexity lower bounds, but mostly using the minimax framework, and with restrictive assumptions on the systems. For example in \cite{simchowitz:2018a}, the authors derive lower bounds for so-called scaled orthogonal systems (refer to Section \ref{sec:uncontrolled} for a definition). These bounds match ours (up to some constant factors) for this specific class of systems. We observe the same limitations in the other recent papers, see e.g. \cite{sarkar:2018, sarkar:2019, faradonbeh:2018:c2, oymak:2018}.   

There have been interesting studies of the sample complexity of LTI system identification earlier in the control community. In the series of papers \cite{WeyerFIR:1996, weyerLTI:1999, WeyerARX:2000}, Weyer et al. consider general FIR and ARX systems, and use the empirical prediction error rate as a performance criterion (we consider the estimation error instead). To derive their results, they rely on strong mixing assumptions (that actually limit the generality of the systems considered), and their sample complexity lower bounds are not problem-specific, and difficult to interpret. In \cite{tikku:1993}, the authors consider even more specific settings (a subclass of scalar systems), and derive minimax lower bounds only. It is also worth mentioning that identification sample complexity has been also investigated for continuous-tme systems, see e.g. \cite{kuusela2004learning}. 

Finally, we can mention \cite{raginsky:2010} where the author derives lower bounds of the time required to achieve a particular objective in adaptive control. Specifically, the time required to achieve, in LQG problems, a regret no greater than a fixed fraction of time (linear regret) is investigated. The lower bound is not problem-specific, and not really related to the identification sample complexity.


\section{Problem formulation}

We consider the following LTI system:
\begin{equation}\label{eq:model}
x_{t+1} = A x_{t} + w_{t}, \qquad x_0 = 0, 
\end{equation}
with state $x_t\in \mathbb{R}^d$ at time $t$, and where $A \in \mathbb{R}^{d\times d}$ is initially unknown. $w_0, w_1, \dots$ is a sequence of i.i.d. Gaussian random variables with distribution $\mathcal{N}(0, I_d)$. We observe the realization of the state trajectory, and from these observations, aim at identifying the matrix $A$ as accurately as possible. More formally, let ${\cal F}_t$ denote the $\sigma$-algebra representing the observations gathered up to time $t$, i.e., generated by $(x_1,\ldots,x_t)$. At time $t$, an algorithm outputs a ${\cal F}_t$-measurable estimator $\hat{A}_t$  of $A$. We are interested in deriving tight lower bounds on the sample complexity of such estimation procedure, namely in finding lower bounds on the number of observations required to identify $A$ with prescribed levels of accuracy and confidence. 

We further extend our analysis to controlled LTI systems:
\begin{equation}\label{eq:model2}
x_{t+1} = A x_{t} + Bu_t+ w_{t}, \qquad x_0 = 0, 
\end{equation}
with control input $u_t\in \mathbb{R}^p$ at time $t$, and where $B \in \mathbb{R}^{d\times p}$ is also initially unknown. There, the objective is to identify both $A$ and $B$. The observations up to time $t$ are represented by ${\cal F}_t$, the $\sigma$-algebra generated by $(u_0,x_1,u_1,\ldots,x_t)$. At time $t$, an algorithm outputs ${\cal F}_t$-measurable estimators $\hat{A}_t$ and $\hat{B}_t$ of $A$ and $B$, as well as a ${\cal F}_t$-measurable control input $u_{t}$. Again, we are interested in deriving tight lower bounds on the sample complexity of such estimation procedure.

\section{Sample Complexity for Uncontrolled Systems}\label{sec:uncontrolled}

\subsection{Change-of-measure Argument}

In a change-of-measure argument, we pretend that the observations are generated from a slightly different system, parametrized by matrix $A'\neq A$. We inspect the expected log-likelihood ratio of the observations generated under the original system and its modification. This expected ratio naturally depends, in a increasing manner, on the number of observations, and a lower bound of this number is obtained leveraging the data processing inequality \cite{garivier2018explore}. We make this argument precise below.

Let $A'\neq A$. The log-likelihood ratio of the $t$ first observations under $A$ and $A'$  is defined as:
$$
L_t = \log \Big(\frac{f_A(x_1, \dots, x_t)}{f_{A'}(x_1, \dots, x_t)}\Big),
$$
where $f_A(x_1, \dots, x_t)$ denotes the probability density (w.r.t. Lebesgue measure) of the observations under the system parametrized by $A$. The Markovian nature of the system dynamics yields: 
$
f_A(x_1,\ldots,x_t)=\prod_{s=1}^t f_A(x_t|x_{t-1}),
$
where $f_A(\cdot |x)$ denotes the density of the state distribution at time $t$ given that it was $x$ at time $t-1$. This density is that of the Gaussian distribution $\nu_{A,x} = {\cal N}(Ax,I_d)$. Let $\mathbb{E}_A[\cdot]$ denote the expectation corresponding to the original system parametrized by $A$. We obtain:
\begin{equation*}
\begin{aligned}
\E_A \big[ & L_t \big]  = \E_A \Big[ \sum_{s=1}^t \E_A \big[ \log\Big(  \frac{f_A(x_s \vert x_{s-1}) }{f_{A'}(x_s \vert x_{s-1})}\Big) \vert  \mathcal{F}_{s-1}  \big] \Big]  \\
    & = \E_A \Big[ \sum_{s=1}^t \KL(\nu_{A,\, x_{s-1}}, \nu_{A',\,x_{s-1}})\Big] \\
    & = \E_A \Big[ \sum_{s=0}^{t-1} \frac{1}{2} x_s^\top(A-A')^\top (A-A')x_s \Big]  \\
\end{aligned}    
\end{equation*}
The last equality in the above is obtained by plugging the explicit expression of the KL divergence of two Gaussian distributions. Observe that: 
\begin{align*}
\E_{A}\Big[ \sum_{s=0}^{t-1} & x_{s}^{\top} (A-A')^{\top} (A-A') x_{s} \Big]  \\
&  =   \tr\Big(   (A-A')^{\top}ß(A-A')\sum_{s=0}^{t-1} \E_{A} \big[ x_{s}x_{s}^{\top} \big]  \Big). 
\end{align*}
Further note that: for all $s\ge 1$,
 \begin{align*}
            \E_{A} \big[ x_{s} x_{s}^\top \big] & = \E_{A} \big[ (\sum_{k=0}^{s-1} A^k w_{s-1-k} ) (\sum_{k=0}^{s-1}  w_{s-1-k}^\top (A^k)^\top)\big] \\ 
            & = \sum_{k=0}^{s-1} A^k (A^k)^\top = \Gamma_{s-1}(A). 
 \end{align*}
 $\Gamma_{s-1}(A)$ is the finite-time controllability gramian of the system. Finally, we have proved that:
 \begin{equation}\label{eq:log}
 \E_A \big[ L_t \big]  ={1\over 2}  \tr\Big( (A-A')^\top (A-A') \E_{A} \big[ \sum_{s = 1}^{t - 1} \Gamma_{s-1}(A) \big] \Big).
 \end{equation}
 The data processing inequality \cite{garivier2018explore} yields: 
\begin{equation*}
 \E_A \big[ L_t \big] \ge \sup_{\mathcal{E} \in \mathcal{F}_t} d(\mathbb{P}_{A}(\mathcal{E}), \;\mathbb{P}_{A'}(\mathcal{E}) ), 
\end{equation*}
where 
$d(x,y) := x\log(x/y) + (1-x)\log((1-x)/(1-y))$ is the KL divergence between two Bernoulli distributions of respective means $x$ and $y$, with the convention that $d(0, 0) = d(1,1) = 0$. Combining this with (\ref{eq:log}) leads to the following result:

\medskip

\begin{proposition}\label{prop1}
For any $A\neq A'$, for all $t \ge 1$,  
\begin{align*}
\tr\Big( (A-A')^\top (A-A') \E_{A} & \big[ \sum_{s = 1}^{t - 1} \Gamma_{s-1}(A) \big] \Big)\\
& \ge 2 \sup_{\mathcal{E} \in \mathcal{F}_t} d(\mathbb{P}_{A}(\mathcal{E}), \;\mathbb{P}_{A'}(\mathcal{E}) ).
\end{align*}
\end{proposition}

\subsection{Locally-Stable Algorithms} 

To derive problem-specific sample complexity lower bounds, we need to restrict our attention to algorithms that really adapt to the system we aim at identifying. Indeed, if the system dynamics is driven by $A$, the algorithm outputting $\hat{A}_t=A$ for all $t\ge 1$ is exact for this system but would be inaccurate for all other systems. Such a restriction is not necessary when deriving minimax sample complexity lower bounds. Hence here, we impose that the algorithms considered are {\it locally-stable} in the following sense:
\begin{definition} ({\it $(\epsilon,\delta)$-locally-stable algorithms})
An algorithm is $(\epsilon,\delta)$-locally-stable in $A$ for some $\epsilon>0$ and $\delta \in (0,1)$  if there exists a finite time $\tau$ such that for all $t \ge \tau$ and all $A'\in B(A,3\epsilon)$,
$$
\mathbb{P}_{A'}(\lVert \hat{A}_{t} - A' \rVert_F \le \epsilon) \ge 1 - \delta,
$$
where $\lVert . \rVert_F$ denotes the Frobenius norm and 
$
B(A, 3\epsilon) = \lbrace A' \in \mathbb{R}^{d \times d}: \lVert A - A'\rVert_F \le 3\varepsilon  \rbrace.
$
\end{definition}
  
The sample complexity $\tau_A$ of an algorithm $(\epsilon,\delta)$-locally-stable in $A$ is then defined as the infimum of the number of observations $\tau$ compatible with the above definition. This means that when the number of observations exceeds the sample complexity in $A$, the algorithm outputs an estimator of $A$ $\epsilon$-accurate with probability at least $1-\delta$.

Note that the existence of locally-stable algorithms is a consequence of earlier results on LTI systems, see e.g. \cite{Ljung:1986:SIT:21413, Goodwin:1977:DSI} or \cite{simchowitz:2018a} for upper bounds on $\tau_A$.         

The notion of $(\epsilon,\delta)$-locally-stable algorithms allows us to lower bound $\sup_{\mathcal{E} \in \mathcal{F}_t} d(\mathbb{P}_{A}(\mathcal{E}), \mathbb{P}_{A'}(\mathcal{E}) )$ for $t\ge \tau$, and in turn lower bound the sample complexity in view of Proposition \ref{prop1}. More precisely, consider an $(\epsilon,\delta)$-locally-stable algorithm in $A$. Let $A'$ be an arbitrary matrix satisfying $2\epsilon\le \|A'-A\|_F<3\epsilon$, and $t\ge \tau$. Define the ${\cal F}_t$-measurable event: $\mathcal{E} = \big\{ \lVert A-\hat{A}_{t} \rVert_F \le \epsilon \big\}$. Since the algorithm is locally-stable, we have:
\begin{align*}
            \mathbb{P}_A(\lVert A-\hat{A}_{\tau} \rVert_F \le \varepsilon ) &\ge  1- \delta   \\
            \mathbb{P}_{A'}(\lVert A-\hat{A}_{\tau} \rVert_F \le \varepsilon ) & \le \mathbb{P}_{A'}(\lVert A'-\hat{A}_{\tau} \rVert_F > \varepsilon )  \le \delta.   
        \end{align*}
We deduce that:  
\begin{align*}
d(\mathbb{P}_A(\mathcal{E}), \mathbb{P}_{A'}(\mathcal{E})) &\ge d(1-\delta, \delta)  \\
            & = (2\delta - 1) \log (\frac{1-\delta}{\delta})\\
            & \ge \log(\frac{1}{2.4 \, \delta}). 
\end{align*}
Combining this result with Proposition \ref{prop1} yields:

\medskip

\begin{proposition}\label{prop2}
For any $(\epsilon,\delta)$-locally-stable algorithm in $A$, for all $A'$ such that $2\epsilon\le \|A'-A\|_F<3\epsilon$, and all $t\ge \tau$, 
$$
\tr\Big( (A-A')^\top (A-A') \E_{A} \big[ \sum_{s = 1}^{t - 1} \Gamma_{s-1}(A) \big]  \Big) \ge 2\log(\frac{1}{2.4 \, \delta}).
$$
\end{proposition}

\medskip

\subsection{Optimizing the Lower Bound}

The previous proposition provides a lower bound of the sample complexity $\tau_A$ of any $(\epsilon,\delta)$-locally-stable algorithm in $A$ for each matrix $A'$ such that $2\epsilon\le \|A'-A\|_F<3\epsilon$. We can finally optimize this lower bound by varying $A'$. The best lower bound can be obtained by solving the following optimization problem:
\begin{equation*}
\begin{aligned}
    \min_{M \succeq 0} & \quad \, \tr\Big( \sum_{s=1}^t \Gamma_{s-1}(A) M \Big) \\
    s.t.               & \quad \tr(M) \ge 4 \epsilon^2
    \end{aligned}
    \end{equation*}
where $M$ plays the role of $(A-A')^\top (A-A')$, and the constraint corresponds to the fact that we impose $2\epsilon\le \|A'-A\|_F$. Note that $\sum_{s=1}^{t -1} \Gamma_{s-1}(A)$ is a symmetric definite positive matrix, the solution to this problem can be obtained directly by choosing $M$ as: 
$$
M = 4\epsilon^2 Q E_d Q^T, \qquad E_d = 
 \begin{bmatrix} 
     0 &       &   &  \\
       &\ddots &   &  \\
       &       & 0 &  \\
       &       &   & 1
     \end{bmatrix}
$$
where $Q$ is the orthogonal matrix that diagonalizes $\sum_{s=1}^{t -1} \Gamma_{s-1}(A)$, placing its eigenvalues in decreasing order. It is important to observe that there exists a matrix $A'$ corresponding to this choice of $M$ and satisfying $2\epsilon\le \|A'-A\|_F<3\epsilon$. Thus we have established the following theorem.

\medskip

\begin{theorem}\label{th:uncontrolled}
For any matrix $A$, for all $\epsilon>0$, $\delta \in (0,1)$, the sample complexity $\tau_A$ of any $(\epsilon,\delta)$-locally-stable algorithm in $A$ satisfies:
$$
\lambda_{\min}\Big(\sum_{s=1}^{\tau_A -1} \Gamma_{s-1}(A) \Big) \ge \frac{1}{2\epsilon^2}\log(\frac{1}{2.4 \delta})
$$
where $\lambda_{\min}(S)$ of some symmetric matrix $S$ corresponds to its minimum eigenvalue.
\end{theorem}

It is worth mentioning that the theorem holds (with the same lower bound) if the performance guarantee for the estimator $\hat{A}_t$ of $A$ is defined through $\| \hat{A}_t-A\|_2$ (instead of the Frobenius norm). To see that, we just need to change the Frobenius norm to the operator norm in the  definition of locally-stable algorithms, and to replace, in the above optimization problem, the constraint $\tr(M)\ge 4\epsilon^2$ by $\sigma_{\max}(M)\ge 4\epsilon^2$ (a constraint on the maximal singular value of $M$). The new optimization problem has the same solution as that of the initial problem.

\subsection{Simplified Lower Bounds}

The lower bound derived in Theorem \ref{th:uncontrolled} depends on the spectrum of the finite-time controllability gramians. Next we provide a looser but more explicit bound. To this aim, we re-start our analysis from the result of Proposition \ref{prop2}. 

Denote by $\lambda_1(A),\ldots,\lambda_d(A)$ the complex eigenvalues of $A$ ordered in decreasing order of their amplitude, i.e., $|\lambda_1(A)| \ge \ldots \ge |\lambda_d(A)|$. The Schur decomposition of $A$ is:
$$
A = Q U Q^\top, \qquad \textrm{where} \qquad
U = \begin{bmatrix}
\begin{matrix}
 B_1 & \times \\
      & B_2
\end{matrix} & 
\begin{matrix}
 \dots & \times \\
  &     
\end{matrix} \\
\begin{matrix}
O
\end{matrix} & 
\begin{matrix}
\ddots     & \times\\
           & B_k 
\end{matrix}
\end{bmatrix},
$$
where the blocks $B_1, \, \dots, B_k$ are either $1 \times 1$ matrices corresponding to the real eigenvalues of $A$ or $2 \times 2$ matrices representing  pairs of complex conjugate eigenvalues of $A$. The orthogonal matrix $Q$ is chosen such that the block $B_k$ corresponds to the eigenvalue(s) of $A$ with smallest amplitude, $\lambda_d(A)$.

In Proposition \ref{prop2}, we then choose $A' = A - 2 \epsilon E_{k}(B_k)Q^\top$, where $E_{k}(B_k)$ is a diagonal-by-block matrix where all the blocks are 0 except for the last one corresponding to $B_k$ in the Schur decomposition of $A$. This last block $J(B_k)$ is just 1 when $B_k$ is a $1\times 1$ block. Otherwise, if $B_k$ is a $2\times 2$ matrix, we let $J(B_k) = P^{-1}\lVert P^{-1}\rVert_F^{-1}$
where $P$ is defined so that 
$$B_k = P \begin{bmatrix}
\alpha & -\beta \\
\beta & \alpha
\end{bmatrix} P^{-1} = \lvert\lambda_d(A) \rvert PR_2(\theta) P^{-1}$$
and where $R_2(\theta)$ is a $2\times 2$ rotation matrix.
In this case we have 
$$
A - A' = 2\epsilon \, \lVert P^{-1} \rVert_F^{-1} \,\begin{bmatrix}
\begin{matrix}
 0 &   \\
      & \ddots
\end{matrix} & 
\begin{matrix}
\end{matrix} \\
\begin{matrix}
\end{matrix} & 
\begin{matrix}
0     &   \\
           & P^{-1} 
\end{matrix}
\end{bmatrix} Q^T.
$$
Note that $\lVert A-A' \rVert_F \ge 2\epsilon$. Now, we can compute\\ $\tr\Big( (A-A')^\top (A-A') \E_{A} \big[ \sum_{s = 1}^{t - 1} \Gamma_{s-1}(A) \big] \Big)$ as
\begin{align*}
& \tr\Big( (A-A')^\top (A-A') \E_{A} \big[ \sum_{s = 1}^{t - 1} \Gamma_{s-1}(A) \big] \Big)\\
& = \tr\Big( Q^\top (A-A')^\top (A-A') Q \;\sum_{s = 1}^{t - 1} \sum_{j = 0}^{s-1} U^j (U^\top)^j   \Big) \\
& = 4\epsilon^2   \sum_{s = 1}^{t -1} \sum_{j = 0}^{s-1} \tr \Big( E_k(B_k)^\top E_{k}(B_k) U^j (U^\top)^j  \Big) \\
& = 4\epsilon^2 \lVert P^{-1} \rVert^{-2}_F  \sum_{s = 1}^{t -1} \sum_{j = 0}^{s-1} \lvert \lambda_{d}(A) \rvert^{2j} \\
& \ \ \ \ \ \ \ \ \ \ \times \tr \Big( R_2(j\theta) P^{-1} (P^{-1})^\top R_2(-j\theta)  \Big) , \\
& = 4\epsilon^2 \phi_{\lvert \lambda_d(A)\rvert }(t), 
\end{align*}
where for any real number $a\ge 0$, the mapping $\phi_a:\mathbb{N}\setminus\{0\} \to \mathbb{R}$ is defined as: for all $t\ge 1$,
$$
 \phi_a(t) = \sum_{s=1}^{t - 1} \sum_{k=0}^{s-1} a^{2k} = \begin{cases} 
      t - 1 & \hbox{if} \quad a = 0, \\
      \frac{a^{2t} + t (1- a^2) -1}{(1-a^ 2)^2} & \hbox{if} \quad a \neq 1,\\
      \frac{t(t -1)}{2} & \hbox{if} \quad a=1. \\
   \end{cases}
$$
When the block $B_k$ is $1\times 1$, the same equality can be proven in a similar way. Applying Proposition \ref{prop2} with this choice of matrix $A'$, we obtain:

\medskip

\begin{proposition}\label{generalcase,betterbound}
For any matrix $A$, for all $\epsilon>0$, $\delta \in (0,1)$, the sample complexity $\tau_a$ of any $(\epsilon,\delta)$-locally-stable algorithm in $A$ satisfies:
$$   
\phi_{\lvert\lambda_d(A)\rvert}(\tau_A) \ge \frac{1}{2\epsilon^2}  \log(\frac{1}{2.4 \delta}),
$$
where recall that $\lambda_d(A)$ is the complex eigenvalue of $A$ with smallest amplitude.
\end{proposition}

\subsection{Examples}

We conclude the analysis of uncontrolled systems by exploring two simple cases, namely scalar systems, and those driven by an orthogonal matrix $A$.

\medskip
\noindent
{\bf Scalar systems.} Consider the scalar system $x_{t+1}=ax_{t}+w_t$. Proposition \ref{prop2} implies that for any $(\epsilon,\delta)$-locally-stable algorithm in $a$, and for any $a'$ such that $2\epsilon \le |a'-a|\le 3\epsilon$, we have:
$$
(a'-a)^2\phi_a(\tau_a) \ge 2\log({1\over 2.4 \delta}).
$$
Simply minimizing the l.h.s. over all allowed $a'$ in the previous inequality leads to the simple sample complexity lower bound: 

\medskip

\begin{corollary}
For any scalar $a \in \mathbb{R}$, for all $\epsilon>0$, $\delta \in (0,1)$, the sample complexity $\tau_a$ of any $(\epsilon,\delta)$-locally-stable algorithm in $a$ satisfies:
$$
\phi_a(\tau_a) \ge \frac{1}{2\epsilon^2} \log(\frac{1}{2.4 \delta}).
$$
\end{corollary}

\medskip

The above lower bound matches that derived in \cite{simchowitz:2018a}, and is tight.

\medskip
\noindent
{\bf Scaled orthogonal systems.} Such systems are driven by a matrix $A$ that can be expressed as $\rho O$ where $O$ is an orthogonal  matrix and $\rho \in \mathbb{R}$. In this case, the gramians and their spectra can be easily computed, and from Theorem \ref{th:uncontrolled}, we get:  

\medskip

\begin{corollary}
For any scalar $\rho \in \mathbb{R}$ and any unitary matrix $O$, for all $\epsilon>0$, $\delta \in (0,1)$, the sample complexity $\tau_{\rho O}$ of any $(\epsilon,\delta)$-locally-stable algorithm in $\rho O$ satisfies:
$$
\phi_{\rho}(\tau_{\rho O}) \ge \frac{1}{2\epsilon^2} \log(\frac{1}{2.4 \delta}).
$$
\end{corollary}

\medskip

Again our lower bound matches that derived in \cite{simchowitz:2018a}, up to constant factors, and are tight (at least for stable systems).

\section{Sample Complexity for Controlled Systems }

We now turn our attention to the controlled linear system (\ref{eq:model2}). We briefly outline the analysis, since it is similar to that of uncontrolled systems. In what follows, we assume that the control policy is fixed, i.e., the input control $u_t$ is ${\cal F}_t$-measurable where we recall that ${\cal F}_t=\sigma(u_0,x_1,u_1,\ldots,x_t)$. The change-of-measure argument involves two matrices $A'$ and $B'$ with $(A',B')\neq (A,B)$. We investigate the same log-likelihood ratio as in the uncontrolled case:
$$
L_t = \log \Big(\frac{f_{A,B}(x_1, \dots, x_t)}{f_{A',B'}(x_1, \dots, x_t)}\Big),
$$
where $f_{A,B}(x_1, \dots, x_t)$ denotes the probability density (w.r.t. Lebesgue measure) of the observations under the system parametrized by $A$ and $B$. The expectation of $L_t$ is given by:
\begin{align*}
\mathbb{E}_{A,B}[L_t] = & {1\over 2}\tr\Big(  \begin{bmatrix} (A-A')^\top \\ (B-B')^\top \end{bmatrix}  
  \begin{bmatrix} (A-A') & (B-B')\end{bmatrix} \\
  &\times \E _{A,B} \Big[ \sum_{s=0}^{t -1}  \begin{bmatrix}
x_{s} \\
u_{s}
\end{bmatrix} \begin{bmatrix}
x_{s}^\top & u_{t}^\top
\end{bmatrix}\Big]  \Big),
\end{align*}
where $\mathbb{E}_{A,B}[\cdot]$ denotes the expectation under system parameters $A$ and $B$.
The notion of locally-stable algorithm is updated as follows.

\medskip

\begin{definition} ({\it $(\epsilon,\delta)$-locally-stable algorithms})
An algorithm is $(\epsilon,\delta)$-locally-stable in $(A,B)$ for some $\epsilon>0$ and $\delta \in (0,1)$ if there exists a finite time $\tau$ such that for all $t \ge \tau$ and all $A'$ and $B'$ such that $\| [A\ B]-[A'\ B']\|_F\le 3\epsilon$,
$$
\mathbb{P}_{A',B'}(\lVert [\hat{A}_{t}\ \hat{B}_t] - [A'\ B'] \rVert_F \le \epsilon) \ge 1 - \delta.
$$
\end{definition}

\medskip
As before, we define the sample complexity $\tau_{A,B}$ of an $(\epsilon,\delta)$-locally-stable algorithm in $(A,B)$ as the smallest $\tau$ compatible with the above definition. We can then establish that under any $(\epsilon,\delta)$-locally-stable algorithm in $(A,B)$, we have: for any $A'$ and $B'$ such that $$2\epsilon\le \lVert \begin{bmatrix} A & B \end{bmatrix} - \begin{bmatrix} A' & B' \end{bmatrix}\rVert_F\le 3\epsilon,$$
and for any $t\ge \tau$,
\begin{align*}
\tr\Big( &  \begin{bmatrix} (A-A')^\top \\ (B-B')^\top \end{bmatrix}  
  \begin{bmatrix} (A-A') & (B-B')\end{bmatrix} \\
  &\times \E _{A,B} \Big[ \sum_{s=0}^{t -1}  \begin{bmatrix}
x_{s} \\
u_{s}
\end{bmatrix} \begin{bmatrix}
x_{s}^\top & u_{t}^\top
\end{bmatrix}\Big]  \Big) \ge 2\log({1\over 2.4\delta}).
\end{align*}
Optimizing over the possible matrices $A'$ and $B'$, we establish the following theorem.

\begin{theorem}\label{th:controlled}
For any matrices $A$ and $B$, for all $\epsilon> 0$ and $\delta\in (0,1)$, the sample complexity $\tau_{A,B}$ of any $(\epsilon,\delta)$-locally-stable algorithm in $(A,B)$ satisfies:
$$
\lambda_{\min}\Big(\E _{A,\,B} \Big[\, \sum_{t=0}^{\tau_{A,B} -1}  \begin{bmatrix}
x_{t} \\
u_{t}
\end{bmatrix} \begin{bmatrix}
x_{t}^\top & u_{t}^\top
\end{bmatrix}\Big] \Big)\ge \frac{1}{2 \varepsilon^2} \log(\frac{1}{2.4\, \delta}).
$$
\end{theorem}

\medskip

The above theorem is less informative than its analog for uncontrolled systems, as the lower bound still involves the states and input controls. The lower bound is hard to simplify. However, it suggests that a control policy leading to minimal sample complexity is obtained by solving the following optimal control problem (here with time horizon $T$):
\begin{align*}
&\sup_{u_t\in{\cal F}_t, \forall t\le T}  \lambda_{\min}\Big(\E _{A,\,B} \Big[\, \sum_{t=0}^{T -1}  \begin{bmatrix}
x_{t} \\
u_{t}
\end{bmatrix} \begin{bmatrix}
x_{t}^\top & u_{t}^\top
\end{bmatrix}\Big] \Big)\\
&\hbox{s.t. } x_{t+1}=Ax_t+Bu_t+w_t, \ \ \forall t\le T.
\end{align*}

In the case of scalar systems, the picture is clearer as discussed below. 

\subsection{Scalar case}

Consider the scalar system: $x_{t+1}=ax_t+bu_t+w_t$. From the above analysis, we deduce that under any $(\epsilon,\delta)$-locally-stable algorithm in $(a,b)$, we have: for any $a'$ and $b'$ such that $$2\epsilon\le \lVert (a',b')-(a,b)\rVert_2\le 3\epsilon,$$ and for any $t\ge \tau$,
$$ 
\E_{a,b}\Big[  \sum_{s=0}^{t-1} ((a-a')x_s + (b-b')u_s)^2  \Big]   \ge  2 \log(\frac{1}{2.4\delta}).
$$
Intuitively, the above inequality provides a lower bound on the number of observations required to distinguish the dynamics of systems either driven by $(a,b)$ or $(a',b')$. It also suggests that when the input control is limited in amplitude, i.e., $u_t\in [-u,u]$, then a control policy minimizing the time it takes to distinguish these two hypotheses is constant and consists in always selecting a control input with maximal amplitude. Next we specify the sample complexity lower bound in case of constant control, $u_t=u$ for any $t$. In case of constant control, Theorem \ref{th:controlled} implies that the sample complexity $\tau_{a,b}$ satisfies:
\begin{align*}
\lambda_{\min} \Big( & \begin{bmatrix} \varphi_{\tau_{a,b}}(a) + \phi_{\tau_{a,b}}(a,b) u^2 &  \psi_{\tau_{a,b}}(a,b) u^2 \\
\psi_{\tau_{a,b}}(a,b) u^2 & (\tau_{a,b} -1)u^2 \end{bmatrix}\Big) \\
&\ge \frac{1}{2 \epsilon^2} \log(\frac{1}{2.4\delta})
\end{align*}
where we have defined
\begin{align*}
    \varphi_{\tau}(a) & = \sum_{t=1}^{\tau-1} \sum_{k=0}^{t-1} a^{2k}, \\
    \phi_{\tau}(a,b) & = \sum_{t=1}^{\tau-1} \big(\sum_{k=0}^{t-1} a^{k}b \big)^2, \\
    \psi_{\tau}(a,b) & = \sum_{t=1}^{\tau-1} \big(\sum_{k=0}^{t-1} a^{k}b \big). 
\end{align*}
We can further show that
$$
f_{ a, b, \tau}(u) = \lambda_{min} \Big(\begin{bmatrix} \varphi_{\tau}(a) + \phi_{\tau}(a,b) u^2 &  \psi_{\tau}(a,b) u^2 \\
\psi_{\tau}(a,b) u^2 & (\tau -1)u^2 \end{bmatrix}\Big),
$$
where
\begin{align*}
f_{ a, b, \tau} & (u)  = \frac{1}{2} \Big( \varphi_{\tau}(a) + (\phi_{\tau}(a,b) +\tau - 1)u^2 \\  &-  \sqrt{(\varphi_{\tau}(a) + (\phi_{\tau}(a,b) - \tau + 1)u^2)^2     + 4 \psi_{\tau}(a,b)^2 u^4} \Big).
\end{align*}
Hence the sample complexity satisfies:
$$
f_{a, b, \tau_{a,b}}(u)\ge\frac{1}{2 \epsilon^2} \log(\frac{1}{2.4\delta}).
$$

\addtolength{\textheight}{-3cm}   


\section{Concluding remarks}

In this paper, we have derived sample complexity lower bounds for arbitrary linear systems. These bounds depend in a simple manner on the system parameters for uncontrolled systems, but are less explicit in the case of controlled systems. We hope to further simplify our bounds for the latter systems. We also plan to investigate in more detail how these bounds and their derivation provide insights into  the design of optimal control policies (by optimal we mean returning accurate estimators with high confidence as early as possible).


\bibliography{references}{}
\bibliographystyle{IEEEtran}

\end{document}